# The Value and Use of Data in Chemical Engineering Practice

**Suparna Samavedham[1] and S. Lakshminarayanan[2,3,†]**

[1]Department of Physics, Boston University, Boston, MA 02215, USA, E-mail: suparnas@bu.edu

[2]Department of Chemical and Biomolecular Engineering, National University of Singapore, Singapore 117585, E-mail: laksh@nus.edu.sg

[3]Department of Chemical and Biochemical Engineering, Western University, Ontario, Canada, N6A 5B9, E-mail: lsamaved@uwo.ca

†Corresponding Author

**Abstract**

The ability to generate, organize, analyze, understand and leverage data for sound decision making is a central activity of chemical engineers. Chemical engineers are responsible for the safe, profitable and environmentally friendly operation of chemical facilities – so they are expected to be good at designing and operating chemical processes. To this end, they make use of models which involves planning and conducting experiments in the laboratory or a pilot plant, analyzing the generated data and making use of it for designing large scale industrial systems. In an operating plant, they need to analyze data so as to achieve maximum efficiencies, reduce the use of precious natural resources, minimize environmental degradation, keep the plant safe as well as to help generate value for the customers and stakeholders. This chapter provides a non-technical view of how chemical engineers translate data to the best possible decisions that result in reliable, safe and profitable process design and operations.

**Keywords**: Design of Experiments, Accelerated Discovery, Fault Detection and Isolation, Mathematical Modeling, Optimization, Soft Sensors

## 1. Introduction

In 1883, Lord Kelvin famously said "I often say that when you can measure what you are speaking about, and express it in numbers, you know something about it; but when you cannot measure it, when you cannot express it in numbers, your knowledge is of a meagre and unsatisfactory kind; it may be the beginning of knowledge, but you have scarcely in your thoughts advanced to the state of Science, whatever the matter may be." In a broad context, he is definitely right but we now know that measurements alone are not enough. On the downstream side of measurements, being able to understand the measured data so that we translate it into meaningful actions in order to can create value to the environment, society and stakeholders is critically important. On the upstream side of data collection, one must know the purpose of the scientific endeavour (i.e., know the purpose of measurement) in order to design and perform the most appropriate experiments on the system so that the measurements have meaning and serves the intended purpose. It should be understood that since data is the fundamental entity in this whole enterprise, no amount of modeling expertise can compensate for poor quality and quantity of data. That is why the design of appropriate experiments is incredibly vital.

Thus, the data life cycle in activities relevant to a chemical engineer are:

i. design of the best experiments that helps to generate appropriate data that would help us to understand or investigate the phenomena of interest,
ii. analyze the data collected and convert the data into the simplest form of mathematical model(s) (equation(s)), and
iii. use the model as the basis for action to design new processes or improve existing processes and collect performance data to verify the correctness of our model and decisions.

Steps (i) to (iii) are iterated a few times until we are satisfied with the performance of the process. In fact, the performance data is collected continuously and whenever a large deviation is noticed between expected and actual performance another investigation involving steps (i) to (iii) is initiated. In all these, data is the central and critical resource and it is therefore important for all chemical engineers to gain skills in all the aforementioned areas.

Measurements and models are central to sense making and operating in the real world for not only engineers but to each one of us. Measurements (evidence/data) primarily serve to construct, de-construct and update models. Some models are left as abstractions in our mind (mental models), some are physically constructed (e.g. in a field like civil engineering or architecture), some are put down as text descriptions with mention of relationships between the different players/actors within the system and some models are specified as mathematical relationships between the factors (variables) in the system. Whatever be the type of model, we put it to use in situations where we have to act to bring about a societal or personal benefit. Amongst these myriad forms of models, mathematical models are of great value for engineers as they can be deployed in tandem with powerful computers and data processing algorithms to enable rapid scientific discoveries, develop deep understanding to innovate further and to design best operational strategies.

Before venturing further, it must be stated that *contextually relevant data* is a critical asset for chemical engineers (Munoz & MacGregor, 2016). Just massive quantities of data won't help us very much as much as a decent quantity of informative data does. Thus, the chemical engineer must work with the information technology (IT) and computer infrastructure personnel within the company to organize data in a contextually correct manner so that operationally relevant information can be extracted. Considerable investment and effort is needed to organize, store and retrieve data for analysis purposes. We do not discuss the issues related to data organization and storage but it suffices to say that alignment of sample times is a critical element of data organization and the compression method adopted to store data is to be carefully decided because the accuracy of the analysis depends on the quality of data retrieved from the data vault.

In this chapter, will discuss some of the intellectual pursuits and activities that a chemical engineer is typically involved in and how data is generated and used in the different scenarios. Without getting into specific techniques, we will discuss the impact of data quality on model construction and subsequent use of the model. We will discuss the different model types (static/dynamic, linear/non-linear, lumped/distributed) and elaborate on how the different models are deployed in applications that serve to keep the environment safe and sustainable while keeping the companies profitable. We will take the readers through the entire life-cycle of data from data generation, converting data to information (models), deploying models as part of an operational strategy or policy, to evaluating the outcomes and finally back to data to facilitate continual improvement. A very good read for chemical engineers to get an overview of the past, present and future of *useful data* and *data use* within the broader Artificial Intelligence and Machine Learning framework can be found in Venkatasubramanian (2019).

## 2. Mathematical Models

Chemical engineers generate and use data across multiple scales - from laboratory to pilot plants to large scale industrial settings in order to construct mathematical models to understand, discover, operate, regulate or optimize the processes. However, they do not use data directly but convert it into a mathematical model, namely a set of equations that capture the essence and patterns/trends in the data as discussed below.

In chemical engineering systems, there are fundamental laws that must be obeyed - these are the *conservation of mass, energy and momentum*. These concepts are covered in the first and second years of a typical undergraduate chemical engineering curriculum (Rice & Do, 1995). In addition to these conservation principles, additional relationships called *constitutive equations* deduced from data are required to complete the model. To build such *mechanistic or first principles* models, the engineer must have some understanding of the physico-chemical-biological mechanisms that govern the process. Typically, this is a difficult and long drawn process; however, when completed, such models can serve as a great proxy for the real process and can be exploited for optimal design and operation of the process.

Alternatively, chemical engineers use experimental data to construct empirical data models ranging from simple linear regression models to the now wildly popular deep neural networks, which are relatively easy to develop (given sufficient quality data) but are usually limited in the range of their applicability as compared to the first principles based models. Increasingly popular is another class of models – hybrid models – which incorporate mechanistic approach (for phenomena that has been understood at a fundamental level) and empirical approach (for phenomena that is poorly understood as of now).

Consider a tank receiving water at a flowrate of $Q_1$ (m³/s). Water leaves from the tank at a flowrate of $Q_2$ (m³/s). The tank has a uniform cross sectional area A (m²).

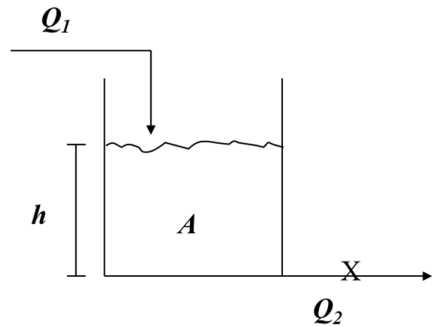

**Figure 1: Schematic of a Water Tank**

Assuming no temperature and composition changes within the tank, the dynamical equation for the volume of water V in the tank can be written as:

$$\frac{dV}{dt} = Q_1 - Q_2 \quad \text{(Eq. 1)}$$

This is essentially the statement of conservation of mass. The water that comes in must accumulate within the tank or leave the tank. At steady state, there is no accumulation within the tank and the volume remains constant. This means dV/dt = 0 and we have the steady state mass balance equation i.e., $Q_1 = Q_2$.

Returning back to the dynamic mass balance equation, we note that since the tank has a constant cross-sectional area, we can write the volume of water (V) in the tank as V = Ah where h is the height of water in the tank. So, Eq(1) may be rewritten as:

$$\frac{d(Ah)}{dt} = Q_1 - Q_2 \quad \text{(Eq. 2)}$$

Since A is a constant, we can write equation 2 as

$$A\frac{dh}{dt} = Q_1 - Q_2 \quad \text{(Eq. 3)}$$

Now, since the flow out of the tank is gravity driven through a pipe at the outlet of the tank, $Q_2$ is related to the value of h. It is understandable that a higher h (hydrostatic head) will lead to a higher $Q_2$ and a low h will lead to a lower value of $Q_2$. Experimental data is collected to establish this relationship. An experimenter sets the liquid level at different h values and measures the $Q_2$ values to establish a mathematical relationship between them. If the relationship turns out to be

$$Q_2 = k\,h \quad \text{(Eq. 4)}$$

we have a linear head-flow relationship.

If the relationship turns out to be

$$Q_2 = k\,\sqrt{h} \quad \text{(Eq. 5)}$$

we have a non-linear head-flow relationship.

Here, equation (4) is a linear algebraic equation while equation (5) is a nonlinear algebraic equation. The parameter k is estimated from experimental data.

If the head-flow relationship is linear, we can put equation (4) into equation (3) to get

$$A\frac{dh}{dt} = Q_1 - k\,h \quad \text{(Eq. 6)}$$

which is a *linear, ordinary differential equation*.

If the head-flow relationship is nonlinear, we can put equation (5) into equation (3) to get

$$A\frac{dh}{dt} = Q_1 - k\,\sqrt{h} \quad \text{(Eq. 7)}$$

which is a *nonlinear, ordinary differential equation*.

Now, equation (6) or (7) can be solved (by hand or computer) to determine how h changes with time t as the inlet flow rate $Q_1$ keeps changing with time.

What we have just demonstrated is first principles modelling where we invoke the law of conservation of mass to come up with the model. In other processes, chemical engineers invoke conservation of energy and momentum to write the governing equations. Also, even in this simple exercise, note that the model could be completed only by determining, experimentally, a relationship between liquid height h and the outflow rate $Q_{out}$. Thus, a model is typically a mix of first principles knowledge and experimentally derived components and the parameters of the model are estimated from experimental/observational data.

All the above types of models are valuable for the practicing engineer to codify their understanding of processes into the most *appropriate* and *simplest* model structures. The word *appropriate* is key because the nature of the system is important in deciding the model structure: Is the process at steady state (i.e., no accumulation of mass or energy within the system) or in a transient state? Is the process behaviour linear or nonlinear? Do the variables depend only on time or position or both or neither? Depending on the answers to these questions, the appropriate form of the model, its structure, is chosen. If the variables do not change with position (e.g. a well-mixed continuous stirred tank reactor), then we have a *lumped model* because the system is homogeneous and can be "lumped". If the variables change with position (as in a plug or tubular flow reactor), then we have a *distributed parameter model*. While lumped parameter models are often described by ordinary differential equations, the distributed parameter models are described by partial differential equations. If variables of a system do not change with time, then a *static model* suffices to capture the system behaviour but if variables change with time, we need a *dynamic model* to capture its behaviour. If the system has characteristics like proportional and symmetric response over the range of interest, then its behaviour is linear and a linear model structure suffices; otherwise a nonlinear model is needed.

Also, in our choice of model structure, we need to ensure that the model does not contain terms that add little or no value to the understanding of the underlying behaviour of the system or to the predictive capability of the model - i.e., the model should conform to the *principle of parsimony*. As a simple example, if we are seeking a model to relate a variable x (say temperature of a liquid stream) to a variable y (say composition of a particular chemical species in the liquid stream) and it is believed that the relationship is nonlinear. Would a quadratic model of the form $y = a_0 + a_1 x + a_2 x^2$ suffice? Or should a cubic term be included i.e., should a model of the form $y = a_0 + a_1 x + a_2 x^2 + a_3 x^3$ be employed? According to the principle of parsimony, the more complex cubic model should be chosen only when it offers a substantial advantage over the simpler quadratic model in predicting y using x values. Chemical engineering students must get enough practice in proposing appropriate model structures and in evaluating them so that they end up with the simplest and most suitable model that they can deploy for the intended purpose. This is often accomplished via a course in applied statistics (Box et al., 2005), typically completed sometime in their sophomore year.

Figure 2 provides an overview of the data generation and use scenarios that a chemical engineering student needs to be introduced to. Firstly, data can be generated by the engineer by performing experiments on the process. Any process can be represented as a system with input variables and output variables. Input variables are typically flow rates, temperature and composition of inlet streams that are processed within an equipment to create products. Output variables are typically the temperature and composition of the product exit streams. Inputs are further classified as manipulated variables and disturbances. Manipulated variables are those that can be set at desired values by the process operator or engineer so as to regulate the key output variables at their target values. Typically, flow rates are considered as manipulated variables as they can be maintained at any value using a control valve. Inlet temperatures, compositions are often disturbance variables as the engineer has to make do with what is delivered by the supplier or an upstream unit. In fact, the operational challenge is to maintain the output values at target values by cleverly adjusting the manipulated variables to compensate for disturbances that affect the process. Thus, a chemical engineer can conduct experiments on the process by perturbing the process inputs (manipulated variables) suitably. This is one way of generating data. Sometimes, no special experiments are conducted and routine data is collected. Routine data is typically information poor and not very useful for process understanding or modelling but they may serve other purposes such as performance monitoring of processes and control loops. As indicated in Figure 2, such data is of immense use to chemical engineers for tasks such as estimating model parameters (e.g. the k value in equation 6 or 7), constructing data-based models (e.g. curve fitting that makes use of correlations between variables), assessing the performance of controllers installed in the process (are they delivering their best possible performance) and to flag process/sensor faults when they occur and to troubleshoot processes. All these activities involve the transformation of a mass of data into appropriate mathematical models (compact representation of the data) and using the models to develop optimal operational strategies in both the short and long term.

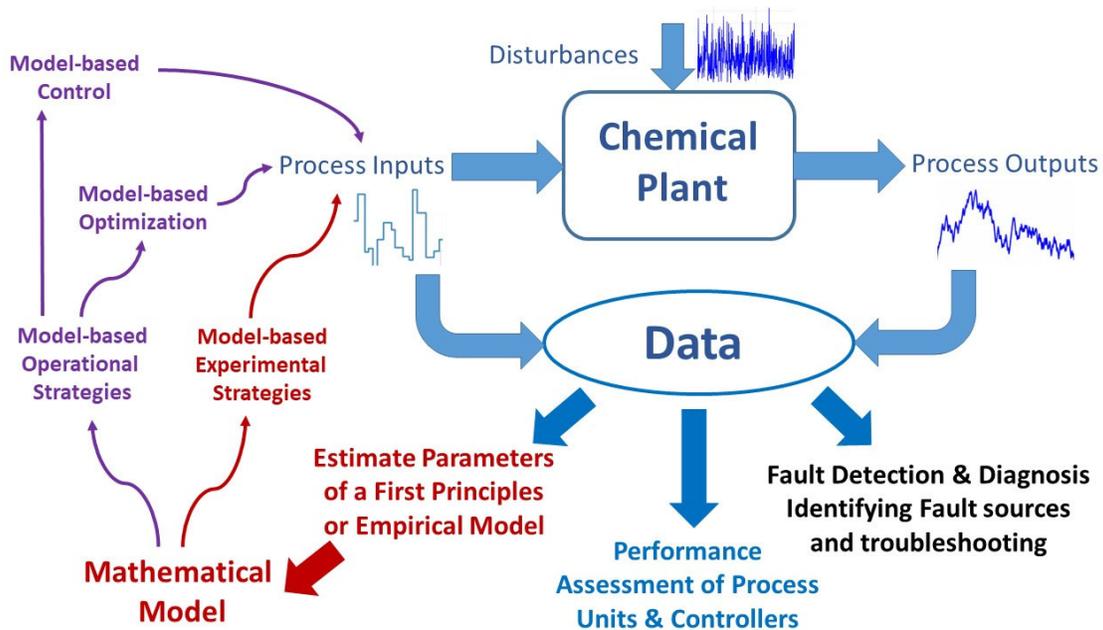

**Figure 2: Generating, Analyzing and Using Data**

## 3. Generating Data: Design of Experiments

Our understanding of the complex, physical-chemical-biological phenomena comes from data that are generated by performing well-planned experiments on the process and elucidating the cause and effect relationships between the variables. For chemical engineers to engage in *sense making* of their processes, the information content in the data (i.e., *data quality*) is perhaps more important than *data quantity*. It is important to for engineering students to understand that millions of data points on thousands of variables can be *non-informative* and no useful information can be obtained from it.

The data quantity requirement depends primarily on the number of variables that needs to be processed and the complexity of the model being developed (e.g. the number of parameters to be estimated in the model). Chemical process data tends to be noisy and while improvements in sensor technology can provide more precise measurements, we will always have to deal with a certain amount of noise (note that any measurement from a sensor can be assumed to be comprised of the true signal value and a noise component; one can envisage additive or multiplicative forms but it suffices to recognize that measurements are always contaminated by noise). Intuitively, we can perhaps accept that the predictive ability of a mathematical model depends on the accuracy of parameter estimates. Mathematically[1], it can be shown that the accuracy of parameter estimates depend on: (i) the number of data samples (quantity), (ii) noise level in the sensor measurements (quality) and (iii) the soundness of the experimental design. The number of data samples is often determined by the time and budget available (note each experiment takes time and costs money in terms of human and material resources). Once the sensors are procured and put in place, one aspect of the data quality (noise in measurements) is pretty much fixed. Fortunately for us, the *soundness of the experimental design* mentioned earlier gives us an opportunity to do the best possible given the time, cost and sensor constraints. This is something that chemical engineers can leverage upon to conduct the *best* experiments to generate better quality data and use it to develop accurate models. This is the critical aspect of the field of design of experiments and something that every experimentalist must be aware of. The key message is that *whenever there is an opportunity, it is best for the chemical engineer to get involved in the data generation (experiment design) process rather than just being content to analyze the data that is passed on by someone*. We might do well to remember the adage *garbage in garbage out* and nothing can be close to it than the activity of modeling with data.

There are two broad approaches to design of *best* experiments - one that is very generic in nature and another that is tailored to the model structure for which we may also know the rough values for some or all of its parameters. The *generic* design of experiments corresponds to designs such as factorial designs, extensions of factorial designs (e.g. as central composite design, Doehlert design, Box Behnken design), fractional factorial designs, and hierarchical/nested experimental designs (Box et al., 2005). Factorial designs and their extensions are simple designs and are applicable in situations where the number of factors that affect the outcome of interest is relatively small. In situations where the number of factors that affect the outcome is significantly high, screening experiments are used to zone in on the most important factors and then factorial design experiments can be performed on this reduced set of variables. Fractional factorial designs are very good screening experiments that can be expanded into full factorial designs if needed. Typically, factorial designs and fractional factorial designs are taught in an undergraduate applied statistics course for chemical engineers employing text books such as Box et al. (2005) and Montgomery (2020). Hierarchical designs and model-based experimental designs are taught as part of undergraduate elective courses or in graduate level courses.

---

[1] Most introductory statistical textbooks provide expressions for the uncertainty in estimated parameters of a linear regression model. That expression will show these three aspects explicitly. Alternatively, look for standard error in parameter estimates which will show (ii) and (iii) explicitly.

The other family of *best* experiments is what is called *model-based design of experiments*. Whatever the origin of the model (first principles or empirical), whatever form it is (algebraic, ordinary differential equation or partial differential equation; linear or nonlinear), as long as we have the current values for the model parameters, the location of the best experiments to do can be determined (Bates & Watts, 2007; Englezos & Kalogerakis, 2001). This usually involves formulating a constrained optimization problem where the objective is usually to minimize the statistical uncertainty in the estimated parameters and the constraints being the current model, the physical/safety limits on the variables and the number of experimental runs that is being planned for. The solution of the optimization problem will be the suggested location of the next set of experiments.

Sometimes, engineers are faced with a different dilemma. There are two or more competing models that perform equally well on historical data and it is desired to choose the best model. While no model can be proven as "true", it is certainly possible to show that some of the models are false in that they cannot produce a certain experimental observations. When faced with the situation of choosing one of the competing models for further use, chemical engineers have to come up with a *discriminating* set of experiments to determine which of the competing models can be *falsified* and removed from further consideration. The model-based design of experiments idea can be extended for such *model-discrimination* studies as well. The interested reader is referred to Englezos and Kalogerakis (2001) and Bates and Watts (2007) for excellent treatments of this topic.

Whether the experiments are designed based on the generic or model-based philosophies, they must be conducted following the general principles of experimental design namely *randomization, replication and blocking*. The recommended experiments are then carried out in a random order (principle of randomization in experimental design) and following strict, consistent protocols (principle of blocking in experimental design). Experiments are replicated (principle of replication in experimental design) if resources and time permit. Without adhering to these principles, the data collected will not have the information content that is required for proper statistical analysis and inference (Box et al., 2005).

## 4. Analyzing Data: Modeling and Parameter Estimation

This is an aspect on which a lot of emphasis is placed in the training of the chemical engineer. This also happens to be the one which is relatively easy particularly if one is not creating new data analysis algorithms but is only applying established methods procedurally albeit in a thoughtful manner. The proliferation of powerful commercial software (e.g. MATLAB, Mathematica, JMP, SPSS) and open source software (e.g. Python, R) has made it easy to build and compare several types of models and their performance and pick the best one for actual deployment. The interested reader can use an efficient search engine and search "software for parameter estimation" to see the number of useful hits it generates. Besides libraries and toolboxes available in the above-mentioned software, the reader might benefit from software such as EASY-FIT (Schittkowski, 2002a) and the related textbook (Schittkowski, 2002b). Python and R are general purpose software that can be used to write codes and libraries for parameter estimation while a software such as EASY-FIT is specialized and tailor-made for chemical and petroleum engineering applications.

The choice of an appropriate model structure is central to the act of modeling and depends on the intended purpose. The intended purpose sets the tone for experiment design (i.e., the best input signal to probe the process and collect the data), the choice of model structure and the type of data that needs to be collected. If the intention is to relate the temperature of the top tray in a distillation column (T) to distillate purity ($X_D$) at steady state, then a linear algebraic model such as $X_D$ = a T + b or a nonlinear algebraic model such as $X_D$ = a T + b $T^2$ + c would be appropriate. In the above expressions, symbols a, b and c would be called the model parameters that must be estimated from plant data. If the relationship between the tray temperature and distillate composition is dynamic in nature (that is, the current value of $X_D$ is determined by the past value of T), then we can propose models of the following form and estimate the parameters of the model from available data:

Linear Dynamic Model: $X_D(t) = a\,T(t-2) + b\,T(t-3) + c$

Nonlinear Dynamic Model: $X_D(t) = \frac{a\,T(t-2)}{1+b\,T(t-2)+c\log(T(t-3))} + d$

If the intention is to develop a linear controller to control the process around a steady state operating point, then a *linear dynamic model* is desired. For this, the best input perturbation to the plant is a pseudorandom binary signal (PRBS), the data on the plant inputs and outputs would be time series data and an appropriate model structure could be a second order plus time delay transfer function or a linear state space model structure. If the idea is to design a nonlinear controller to operate the process at a point where it exhibits severe nonlinearity or over a wide operation zone, then a nonlinear model structure is appropriate and multi-level input signal serves as the best probing signal to excite the process. Again, time series data musty be collected. If the intention is to develop a model that captures the steady state behaviour of a process, then an algebraic model structure is chosen and steady state operating data is collected (time series data is not needed for this case).

With the collected data, the parameters of the chosen model would be determined by minimizing a measure such as the squared prediction error (sum of squared deviations between experimental data and model predictions). Minimizing the squared prediction error is by itself not a good goal as it could be driven smaller by just making the model more complex than required which leads to what is called *overfitting*. So the objective function that is minimized to estimate unknown parameters of the model should be able to prevent *underfitting* and *overfitting*. To do this, the objective function includes a penalty term that is related to model complexity in addition to the term that is related to the squared prediction error. When more parameters than is necessary gets included in the model, it reduces the squared prediction error but increases the complexity term in the objective function. When less than necessary parameters are used in the model, the complexity term drops but the squared prediction error becomes high. Thus, it is easy to see that the composite objective function can help find a sweet spot that balances model complexity and its predictive capability. This is formally incorporated in information-theoretic measures such as the Akaike Information Criteria (AIC), Bayes Information Criteria (BIC), and Final Prediction Error (FPE) in modeling software. Prediction Error Methods (PEM) or Maximum Likelihood Methods (MLE) are some of the extensively used model parameter estimation methods for time series data. For algebraic models, linear or nonlinear least squares methods may suffice.

Proposing a suitable model structure and estimating its parameters along with their statistical confidence intervals is a relatively easy task with modern day computers and modeling software once appropriate data is available. Before deploying the model for its intended purpose in actual production units, it is important for the constructed model to be tested against *new* data sets to check if model predictions match well with experimental values. If this *model validation* step is successful, the model can be deployed in applications with confidence. For industrial practitioners, extensive and adequate validation of the model with fresh data sets is the key for implementation. Academics tend to focus a lot on issues such as whether the assumptions made in the model hold true – for example, is the noise of constant variance? is the noise independent and normally distributed? This is called residual analysis or diagnostics and it is important that students are made aware of these issues – checking if model assumptions are valid, monitoring model performance (predictions) continuously etc. to identify sensor drift, changing process behaviour etc.

Now that we have a reasonable overview of the main elements related to process, data and models and the intimate connections between them, we can elaborate on some of the typical applications that a chemical engineer deals with in the course of his/her professional practice.

## 5. Typical Applications

### Application 1: Hypothesis Testing

One very common use of data in chemical plants is in hypothesis testing. A typical situation is like this: Catalyst A is currently employed by the operations team in a chemical reactor and is known to provide an average yield of 70% for a desired product. Meanwhile, a chemist working in the R&D department has invented a new Catalyst B and claims that it provides a higher yield of the desired product than does Catalyst A. Should this claim be accepted and the process operations team adopt Catalyst B? What is done in such cases is that a careful experimental design is conducted (following principles of randomization, blocking and replication) and data is collected for both catalysts. Depending on the characteristics of the experimental design and the number of experimental runs with the two catalysts, an appropriate statistical test (t-Test or Z-test) is conducted to see if there is evidence that the average yield obtained with catalyst B is higher than the average yield obtained with catalyst A. If the answer is yes, then the operations team will adopt catalyst B. The same idea can be extended to situations such as "Does Adsorbent A work better than Absorbent B for $CO_2$ capture?", "Does fermentation Process A or fermentation Process B produce better tasting beer?" or "Do inexperienced operators cause more plant accidents than experienced operators?".

### Application 2: Soft Sensors

Yet another common use of data in chemical engineering practice is the design of soft sensors. In a typical process, it is very easy to get frequent measurements (once every few seconds) of variables such as temperature, pressure, level and flow rates. This is because sensor development for these variables is quite mature and they are less expensive (compared to composition sensors, viscosity meters, pH meters) as well. On the other hand, it is not as easy or cheap to measure product quality variables such as composition, melt flow index, viscosity, number average molecular weight and so on. While inline sensors for these variables is increasingly available, current practice is to draw samples and do offline laboratory analysis using sophisticated analytical equipment and record the values. It is not uncommon for such product quality variables to be measured once every 30 minutes. When product quality measurements are available only infrequently, it is easy to make products that are off-specifications, leading to consumer dissatisfaction because disturbance variables affect the process. It is important to compensate for these disturbances by adjusting some variables (called manipulated variables) within the system and continue to deliver the product quality that was intended. In the absence of hardware sensor-based measurements of the product quality variables, one interesting option is to construct a model between the frequently measured variables and the infrequently measured quality variables by using appropriate sections of past historical data (data selection). The constructed model is referred to as the soft sensor. The "mathematical model" shown in Figure 2 includes such soft sensors and is part of the improved operational strategies.

Soft sensors are heavily deployed in chemical plants, petrochemicals, and biologics manufacture as part of monitoring and inferential control strategies. Yet another area is in using spectral data (FTIR, NMR etc.) to estimate product quality such as purity and solubility in chemical plants and in novel applications in the food industry (Bruwer et al., 2007).

In the early stages of soft sensor development, the ordinary least squares (multiple linear regression, MLR) technique was employed. Later, with more data being available and with the variables being highly correlated with each other, latent variable based techniques like Principal Components Regression (PCR), Partial Least Squares (PLS) etc. came into prominence. In the recent past, machine learning approaches such as support vector regression (SVR), artificial neural networks (ANNs), and their extensions have become popular and offer the ability to capture nonlinear and dynamic relationships. MLR is typically covered in a Year 2 chemical engineering course and other methods are typically taught in elective courses.

Model building is just one part of the story. With powerful algorithms that are easily accessible, it is quite easy to construct soft sensor models. The models have to be validated by checking its predictive capability on data sets that were not used during model construction phase. Only when the model validates well against multiple fresh data sets, then the soft sensor model can be deployed online. Even then, its performance has to be monitored on a frequent basis. If its performance deteriorates, then the model is recalibrated or reconstructed starting the entire cycle of data selection, model construction, model validation and online deployment again

For an in-depth treatment of this topic, the reader is referred to Kano and Fujiwara (2013).

**Application 3: Optimization of Processes**

It is often the case that the performance of a chemical process system needs to be optimized. For example, the conversion of a reactant should be maximized or the number of cracks in the product of a metallurgical process should be minimized. To achieve such optimal outcomes, systems usually have handles which we are able to adjust to our will (of course, within physical and safety constraints). These handles, called decision variables, should causally influence the output of interest and can be set to certain specific values that optimize (maximize or minimize) the outcome of interest. For example, if we want to maximize reactant conversion in an industrial reactor, we may be able to use reactor temperature, reactor pressure and catalyst type to achieve the goal. In this case, the objective function (which needs to be maximized) is conversion of reactant ($X$) and the decision variables are reactor temperature ($T$), reactor pressure ($P$) and catalyst type ($C$). $T$ and $P$ are continuous variables and $C$ is a discrete variable (we may only have three specific catalysts $C_1$, $C_2$ and $C_3$ available). This sets up a constrained optimization problem as follows:

$$\begin{aligned}
\underset{T,P,C}{\text{maximize}} \quad & X \\
\text{subject to} \quad & T_{min} \leq T \leq T_{max}, \\
& P_{min} \leq P \leq P_{max}, \\
& C \in [C_1, C_2, C_3]
\end{aligned}$$

The important element of this problem is to have a model that relates $X$ to $T$, $P$ and $C$. If we are able to write a first principles model relating $T$, $P$ and $C$ to $X$, then we are ready to solve the optimization problem. However, such models are often not available in industrial settings. Engineers then resort to the use of historical data if an information-rich data set is available and construct an empirical model. If such data set is not available, then engineers go about generating such a data set using one of the well-studied experimental designs such as factorial designs and fractional factorial designs. The reader can refer to Box et al. (2005) for these experimental designs and more. These experimental designs provide the location of experimental points that effectively span the available search space (between $T_{min}$ and $T_{max}$ for $T$; between $P_{min}$ and $P_{max}$ for $P$ and from the set {$C_1$, $C_2$ and $C_3$} for $C$). The recommended experiments are then carried out by the engineers adhering to the principles of experimental design mentioned earlier namely randomization, blocking and replication. When the experiments are completed, a suitable model is determined based on the ideas discussed in section 4. In many cases, the model takes the form with linear and interaction terms (motivated by the dominant terms of the Taylor series expansion of well-defined arbitrary functions) as exemplified in the equation below for our example system.

$$X = a_0 + a_1 T + a_2 P + a_3 C + + a_4 T*P + a_5 T*C + a_6 P*C + a_7 T*P*C$$

where $a_0$ through $a_7$ represent the model parameters that are determined from data. Once we have a suitable model relating $T$, $P$ and $C$ to $X$ (obtained from first principles or by the method described above), we are able to solve the optimization problem using a suitable optimization algorithm. Specific optimization methods like Linear Programming (LP), Quadratic Programming (QP), Mixed Integer Linear Programming (MILP) and Mixed Integer Nonlinear Programming (MINLP) are covered in chemical engineering elective courses or at the graduate level. Several textbooks by chemical engineering faculty are available - the beginner may start with the textbook (Edgar et al., 2001).

**Application 4: Accelerated Product Discovery and Design**

One of the things facilitated by data analysis is our ability to rapidly discover products with better or customer-specific physical/chemical/functional properties. For example, a U.K. based AI company DeepMind announced a product AlphaFold that quickly and accurately predicts the three dimensional shape of any protein. While this is spectacular in itself, AlphaFold utilizes protein structures and their folding patterns from large databases and combines it with physical and biological knowledge about protein structures to predict protein structures with atomic accuracy even when no similar structure is known. This development has the potential to impact drug design in pharmaceuticals and to optmize enzymes for use in biologics manufacturing.

Materials discovery for energy, medical and environmental applications involves *ab initio* calculations and molecular dynamics (MD) simulations that are computationally intensive. Therefore, while these simulations offer accurate answers, they become a bottleneck in the discovery of novel materials. Using data from past *ab initio* calculations and MD simulations, simpler models (called *surrogate models*) that have sufficient accuracy but execute very swiftly can be constructed. Use of surrogate models in place of the original models speeds up the exploration of the multi-parametric search space by orders of magnitude (e.g. in designing a novel adsorbent for $CO_2$ capture). Utilizing this idea, significant progress has been made in materials modeling and design to serve a variety of areas including organic chemistry, biomolecules, inorganic crystals to surface science (Gubernatis and Lookman, 2018). The surrogate model based methods will smartly complement traditional simulation methods that are based on classical and quantum mechanics calculations. Fast exploration with surrogate models will help to identify promising material designs which can then be validated by detailed molecular/quantum mechanics calculations. A very recent article that exemplifies the synergistic interplay between molecular dynamics simulation, regression and optimization (described above) to design solvents for crystallization can be found in Chai et al. (2021). Such integrated approaches should be taught to students to prevent a silo-ed approach to learning and applications.

Chemical process and product design depend heavily on the knowledge of physiochemical properties of pure components. Chemical engineers strive to develop models to relate molecular structural information of chemicals to a myriad of pure component properties such as enthalpies of formation, fusion, vaporization, melting and boiling points, critical properties, solubility etc. This can be thought of as a soft sensor model that predicts physiochemical properties from easily obtainable molecular structures. Both traditional (regression) and modern (machine learning) approaches can be utilized for this as presented in Alshehri et al. (2021). These soft sensor models serve as computationally attractive surrogate models for detailed calculations or expensive and time consuming experiments and can therefore considerably speed up the design of processes and products.

**Application 5: Fault Detection and Diagnosis**

With increased emphasis on increased operating efficiencies, stricter environmental regulations, and sustainable use of natural resources, chemical processes are designed with material and energy integration pathways. While this has made the plants more efficient, they have also rendered them more complex from an operations perspective. Disturbances and faults that originate in one part of the system do not propagate and leave the system. They keep circling back and affect many parts of the system including upstream units. Thus process faults and plant upsets are becoming costlier than before and increasingly difficult to locate and ameliorate.

With well instrumented chemical plants spewing out huge volumes of data at an ever increasing velocity, it is but natural to capitalize on this very important asset and develop data-based process monitoring approaches. An industrially successful approach has been to construct a statistical model of the process based on data collected when the plant was subject to *common cause* (*non-systemic*) *variations* only and no customer complaints were registered. This *normal operations* model can then be used to query the current plant status; as long as the plant is subject to common cause variations, the current plant data will not deviate significantly from the *normal operations model*. When *systemic variations* (*assignable cause variations*) strike the process, the variables can be larger or smaller than normal and the relationships between variables can also be altered. When this happens, statistically significant deviations from the *normal operations model* is noticed and the process is flagged as *abnormal* or *out-of-control*. Fault diagnosis procedures are then initiated with a view to pinpoint the source(s) of the problem and make fundamental changes to the way the process is operated following closely the philosophy of statistical process control.

It is very important to be able to distinguish between process changes that are made intentionally (e.g. set point changes, throughput changes and grade transitions) and those that represent real process upsets such as misbehaving or failed sensors/actuators, catalyst deactivation, fouling of heat transfer surfaces etc. There are subtle but important differences in the process dynamics for these two scenarios and this can be exploited by the plant operating personnel.

The increasing interest in methodology development and applications in the process monitoring/fault detection and diagnosis area has resulted in several interesting research and review articles over the past 25 years. Pioneering work in this area (popularly known as latent variable methods) was initiated by MacGregor and co-workers at McMaster University, Canada in the early 1990s (Kresta et al., 1991; MacGregor et al., 1994; Yoon & MacGregor, 2001) and subsequently up by other research groups such as Shah and co-workers (University of Alberta, Canada) (Lakshminarayanan et al., 1996; Li et al., 2003). Parallel contributions were made in the US particularly by Wise and co-workers (e.g. Wise & Gallagher, 1996). These methods employed multivariate statistical techniques such as Principal Components Analysis (PCA) and Partial Least Squares (PLS) to set up multivariate monitoring charts that work reliably even when presented with correlated data sets. Chemical engineering students are taught univariate process monitoring charts such as X-chart, p-chart, CUSUM and EWMA in their undergraduate statistics course. With techniques like PCA and PLS coming to the fore and related software more accessible, chemical engineering curriculum has included them and their applications to the monitoring and fault detection of processes. In these methods, operating data collected under good plant conditions are used to create a small number of orthogonal latent variables that capture the dominant relationships between the measured process variables. These dominant variables constitute the *signal space* and simple ideas from normal probability distribution are used to set statistical limits for the signal space. Whatever is not included in the dominant latent variables is consigned as the *residual space* for which statistical limits can be established. Now, when new plant data is obtained, it is projected against the established statistical limits for both the signal and residual spaces. If the statistical limits for the signal space and/or residual space is violated, the process is deemed out of control and an abnormal status is flagged (fault detection). Simply put, the plant data at each instant or over a time window is subjected to a multivariate hypothesis test that tests the null hypothesis "The process is in statistical control" versus the alternative that it isn't. Once an abnormal status is flagged, the search for the source of the fault (fault diagnosis) is initiated and is considerably more challenging. Many methods that range from the simple *contribution plots* (MacGregor et al., 1994) to much more rigorous methods that include *model-based reasoning* (Rich & Venkatasubramanian, 1987), *syntactic pattern recognition* (Rengaswamy & Venkatasubramanian, 1995) and *dynamic trend analysis* (Maurya et al., 2007) have been proposed and adopted in industrial practice. Use of such data-based monitoring approaches have saved chemical plants, refineries, pulp and paper plant, metal processing plants and pharmaceutical companies hundreds of thousands of dollars in terms of reduced incidents, equipment downtime, need for rework etc. (e.g. Dudzic et al., 2000; Matos et al., 2008).

**Application 6: Performance Assessment of Process Units and Controllers**

It is one thing to design processes and equip them with controllers so as to regulate key variables to keep the processes safe and profitable but it is another thing to evaluate and ensure that the processes are performing to their full potential. For example, while the controllers may be performing well at the time of commissioning, their performance may deteriorate over time because of changes in the process conditions such as fouling of heat exchange surfaces, catalyst deactivation in reactors, fall in tray efficiencies in separation columns etc. The performance of the controllers must be assessed at regular intervals and compared to a benchmark performance. If the performance drops significantly from the benchmark value for some controllers, they may have to be retuned so as to recover the performance. While controller retuning may be easier if experiments are done on the control loops, such experiments may be laborious, disruptive and expensive. The challenge is to use routine operating data and minimal knowledge of the process (only the process delay) to assess the performance of controllers and retune them (Jain & Lakshminarayanan, 2007). Literature and industrial applications in this area started around the mid-1990's with major contributions from Harris and co-workers (Desborough & Harris, 1993), Shah and Huang (e.g. Huang et al., 1997; Huang & Shah, 1998; Huang and Shah, 1999), Tyler and Morari (1996), MacGregor and coworkers (Harris et al., 1996; Stanfelj et al., 1993), Patwardhan and Shah (2002) and the field is considered to be fairly mature now (Jelali, 2006). A process control engineer has to take care of 400 to 500 control loops in a refinery or chemical plant and, to be productive, the engineer must be able to understand how routine closed-loop plant data can be used to assess and improve the performance of controllers.

It is not only about evaluating and maintaining the performance of controllers but also about equipment such as pumps, compressors and valves. If any component of any process unit under-performs, the whole system does not perform to its potential. For example, even with a well-designed process that is regulated by a good controller, the overall performance may be poor if the valve malfunctions. Control valves can develop many problems due to loose or worn mechanical linkages between the elements that make up a valve i.e., positioner, actuator, and/or valve. Such malfunction leads to what can be mathematically be characterized as dead band, backlash and hysteresis. Process operators and engineers continually monitor the valves to spot such issues and troubleshoot them. With the availability of plant data, it becomes relatively easy to analyze the data and quantify the extent of the problem (Choudhury et al., 2005; Choudhury et al., 2010; Jelali & Huang, 2010; Srinivasan et al., 2015; Srinivasan & Rengaswamy, 2008) and to initiate repairs, maintenance or have temporary fixes so that the plant can continue to operate until the next maintenance period (e.g. Lee & Lakshminarayanan, 2009).

Troubleshooting of equipment such as pumps, compressors, industrial boilers, heat exchangers etc. are increasingly being done by using multi-modal data (process variables (e.g. temperatures, pressures, flow rates), measurements from vibration sensors, audio measurements (e.g. recorded noise generated by rotating equipment), images and videos (e.g. flame brightness and shape in a furnace or boiler or at top of exhaust stacks) facilitated by the

*industrial internet of things (IIoT)*. Thus, chemical engineers would find it worthwhile to gain skills in processing a wider range of digital information that comes their way and harness them for better decision making.

**6. Concluding Remarks**

To quote British theorist and academic Stafford Beer "Information is *what changes us*. Up to then the stuff is data. Data is useless, until it is transmogrified into something that changes you". This chapter has hopefully provided some perspective on how chemical engineers think about data and how they use it for improving products and processes that benefit society. Paraphrasing Stafford Beer, we might say that data is useless unless it is converted into information that provides a basis for action to change the processes we operate and the products we manufacture.

Data is beginning to play a greater role in transforming the chemical engineering research and practice together with rapid advances in IIoT, data storage and retrieval, powerful computers, machine learning algorithms and artificial intelligence. Data-based applications will increasingly address harder problems and global challenges related to food, water, health, energy and environment in which chemical engineers will have to make critical contributions. It therefore makes sense to invest in educational programmes and courses wherein chemical engineers are taught to generate, use and reuse data responsibly and productively. It is important that the discussions and training should focus on all aspects of the data cycle i.e., generation, storage, organization, use and reuse of data rather than obsessively focusing only on the analysis of data.

**ABOUT THE AUTHORS**

Suparna Samavedham has a BA in Physics and Mathematics (Statistics) with a minor in Psychology at Boston University, USA. Her interests include Data Science and Analytics as applied to Science and Engineering domains.

Lakshminarayanan Samavedham is an Associate Professor at the Department of Chemical and Biomolecular Engineering, National University of Singapore (NUS) and an Adjunct Research Professor at the Department of Chemical and Biochemical Engineering at Western University, Canada. He has worked in the area of process data analytics for over 25 years and has focused on model development for process control, optimization and fault diagnosis. He has published over 100 peer reviewed research articles in high quality journals and has made over 150 conference presentations including several keynote talks. Laksh is also the Director of the Institute for Applied Learning Sciences and Educational Technology (ALSET) at NUS.

**DECLARATION OF CONTRIBUTIONS**

S. S. contributed to this work through research on various subtopics we touch upon in this work, in writing sections on model building, soft sensors and hypothesis testing. She also offered feedback on the remaining sections of the chapter. S.L. conceptualized the chapter, its structure and wrote all the other parts of the chapter.